\documentclass[12pt]{article}
\usepackage{graphicx}
\usepackage{amssymb}
\usepackage{amsmath}

\begin{document}

\begin{center}
{\large \bf Hydrodynamic Crossovers in Surface-Directed 
Spinodal Decomposition and Surface Enrichment}
\ \\
\ \\
by \\
Prabhat K. Jaiswal$^1$, Sanjay Puri$^1$, and Subir K. Das$^{2}$ \\
$^1$School of Physical Sciences, Jawaharlal Nehru University, New Delhi -- 110067, India. \\
$^2$Theoretical Sciences Unit, Jawaharlal Nehru Centre for Advanced Scientific Research, 
Jakkur, Bangalore -- 560064, India.
\end{center}

\begin{abstract}
We present comprehensive molecular dynamics (MD) results for the kinetics of 
surface-directed spinodal decomposition (SDSD) and surface enrichment (SE) 
in binary mixtures at wetting surfaces. We study the surface morphology and 
the growth dynamics of the wetting and enrichment layers. The growth law for 
the thickness of these layers shows a crossover from a diffusive regime to a 
hydrodynamic regime. We provide phenomenological arguments to understand this
crossover.
\end{abstract}

\newpage
A rich class of physical phenomena occurs when a homogeneous binary ($AB$) mixture 
is placed in contact with a surface ($S$). In many cases, the surface has a 
preferential attraction for one of the components of the mixture (say $A$). If 
the mixture is immiscible, the homogeneous bulk is unstable to phase separation 
and segregates into growing $A$-rich and $B$-rich domains. The surface becomes 
the origin of {\it surface-directed spinodal decomposition} (SDSD) waves which 
propagate into the bulk 
\cite{jnkbw91,kdkb93,lwll09,wcl10,gk03,pf97,sp05,dphb05,yx08,ylx08,bpdh10}. 
The system evolves into a {\it partially wet} (PW) or {\it completely wet} (CW) 
equilibrium morphology, depending upon the relative interaction strengths among 
$A,B$ and $S$ \cite{cahn77,mef84,gennes85,sd88}. In the PW morphology,
the interface between the A-rich and B-rich domains
meets the surface at a contact angle determined by {\it Young's condition} \cite{ty05}.
In the CW morphology, the A-rich phase expels the B-rich phase from the surface,
and the AB interface is parallel to the surface.
On the other hand, if the mixture is miscible, the emergent morphology consists 
of a thin {\it surface enrichment} (SE) layer followed by an extended {\it depletion 
layer} \cite{yl06,jkrss89,gnpl97,wdsch03}. 
The kinetics of SDSD and SE are of great technological and scientific importance, 
and find application in the fabrication of nanostructures, layered materials, 
composites, etc. 

Many experiments on SDSD and SE involve polymer or fluid mixtures, where 
hydrodynamic effects play an important role. For bulk phase-separation kinetics, it 
is well known that hydrodynamics has a drastic effect on the intermediate 
and late stages \cite{bray94,pw09}. 
The coarsening domains with size $L(t)$ show a power-law growth, 
$L(t) \sim t^\phi$, with the exponent $\phi$ changing from $1/3$ (diffusive regime) 
to $1$ (viscous hydrodynamic regime) to $2/3$ (inertial hydrodynamic regime) 
\cite{kcpdb01,wc01,adp10}. 
However, there is no analogous understanding of the 
role of fluid velocity fields in SDSD and SE 
\cite{bpl01,tanaka01}. 
In this paper, we 
present comprehensive molecular dynamics (MD) results for the kinetics of SDSD 
and SE. Our MD simulations clearly 
demonstrate a sharp crossover from a diffusive regime to a viscous hydrodynamic 
regime in both SDSD and SE. These results are of great relevance for experimentalists 
in this area, and provide a framework to understand the observation of 
diverse growth exponents in experiments. 

We consider a binary ($AB$) fluid mixture of point particles confined in a box 
of volume $V = L \times L \times D$. Periodic boundary conditions are applied in 
the $x$ and $y$ directions. An impenetrable surface is present at $z=0$, which 
gives rise to an integrated Lennard-Jones (LJ) potential ($\alpha=A,B$):
\begin{equation}
 u_w(z)=\frac{2\pi n\sigma^3}{3} \left[\frac{2\epsilon_r}{15}{\left(\frac{\sigma}{z^\prime}
\right)}^9-\delta_\alpha\epsilon_a{\left(\frac{\sigma}{z^\prime}\right)}^3\right],
\label{eq1:intlj}
\end{equation}
where $n$ is the fluid density, and $\sigma$ is the LJ diameter. In 
Eq.~(\ref{eq1:intlj}), $\epsilon_r$ and $\epsilon_a$ are the strengths of the 
repulsive and attractive parts of the surface potential. We set $\delta_A=1$ and 
$\delta_B=0$, i.e., $A$ particles are attracted at large distances, whereas $B$ 
particles are only repelled. Further, $z^\prime=z+\sigma/2$ so that the singularity 
of $u_w(z)$ occurs at $z=-\sigma/2$ (outside the box). A similar surface is 
present at $z=D$ with $z^\prime=D+\sigma/2-z$, and $\delta_A=\delta_B=0$, i.e., both 
$A$ and $B$ particles are repelled. The simulation box corresponds to a semi-infinite 
geometry \cite{sp05}: the generalization to any other geometry is straightforward.

The particles in the system interact with LJ potentials ($\alpha=A,B$):
\begin{equation}
 u(r_{ij})=4\epsilon_{\alpha\beta}\left[\left(\frac{\sigma}
{r_{ij}}\right)^{12}-\left(\frac{\sigma}{r_{ij}}\right)^{6}\right],
\label{eq2:lj}
\end{equation}
where $r_{ij}=|\vec{r}_i-\vec{r}_j|$. We set the interaction parameters as 
$\epsilon_{AA}=\epsilon_{BB}=2\epsilon_{AB}=\epsilon$. The bulk phase diagram 
for this potential is well known \cite{dhb03,dhbfs06,dfshb06}. 
We use the truncated LJ potential with $r_c=2.5\sigma$ -- this potential is 
shifted and force-corrected \cite{at87}.
We consider a critical composition with $N_A=N_B=N/2$. The system is a 
high-density incompressible fluid with $n=N/V=1$. The particles have 
equal masses ($m_A=m_B=1$); and we set $\sigma=1,\epsilon=1,k_B=1$ so that 
the MD time unit is $t_0=\sqrt{m\sigma^2/(48\epsilon)}=1/\sqrt{48}$. 

The MD runs were performed using the Verlet velocity algorithm \cite{bc96} 
with a time-step $\varDelta t=0.07$ in MD units. We maintain the temperature 
($T$) via the Nos\'{e}-Hoover thermostat which is known to preserve hydrodynamics 
\cite{bc96,adp10}. The homogeneous initial state for a run was prepared 
by equilibrating the system at high $T$ with periodic boundary conditions 
in all directions. At time $t=0$, the system is quenched to $T<T_c$ for 
SDSD ($T_c\simeq 1.423$) or $T>T_c$ for SE, and the surfaces are introduced 
at $z=0,D$. We can also consider initial conditions where the mixture
at high $T$ has been equilibrated with the surfaces at $t<0$. In that case, a
thin enrichment layer arises at $z=0$ -- this layer vanishes at very high values
of $T$. When the system is quenched, this very thin initial layer rapidly becomes
part of the wetting or enrichment layers which grow from the surface.

\begin{figure}[!htbp]
\centering
\includegraphics*[width=0.95\textwidth]{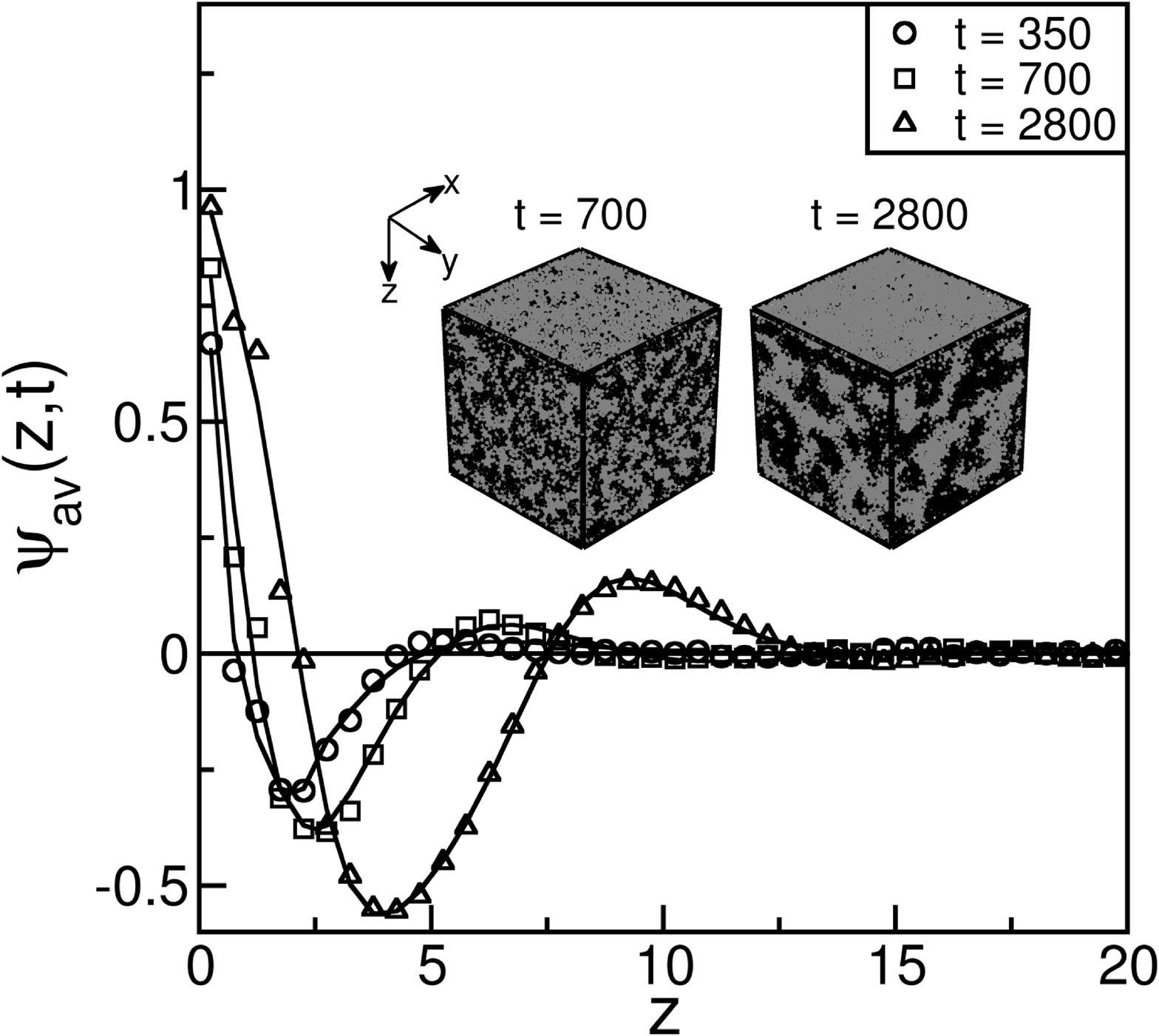}
\caption{Kinetics of surface-directed spinodal decomposition (SDSD) in a binary 
($AB$) Lennard-Jones (LJ) mixture. An impenetrable surface (located at $z=0$) 
attracts the $A$-particles. The surface field strengths are 
$\epsilon_r=0.5,\epsilon_a = 0.6$, and the temperature is $T=1.0\simeq 0.7\,T_c$. 
The other simulation details are given in the text. Main figure: 
Laterally-averaged order parameter profiles [$\psi_{\textrm{av}}(z,t)$ vs. $z$]
at $t=350, 700, 2800$ MD units. 
Inset: Evolution snapshots at $t=700, 2800$. The $A$-particles are marked gray 
and the $B$-particles are marked black.}
\label{fig:fig1}
\end{figure}

First, we present results for the kinetics of SDSD. 
The system size was $L^2\times D$ 
with $L=48, D=48$ ($N=110592$). The surface potential parameters were 
$\epsilon_r=0.5, \epsilon_a=0.6$, and the quench temperature was 
$T=1.0\simeq 0.7T_c$. Equilibrium wetting phenomena for this model
have been studied by Das and Binder \cite{db}, who observed a first-order
wetting transition at the temperatures of interest.
The above parameters correspond to a CW morphology in 
equilibrium. In Fig.~\ref{fig:fig1}, we show the laterally-averaged depth 
profiles [$\psi_{\textrm{av}}(z,t)$ vs. $z$] and the evolution snapshots (inset). 
The snapshot at $t=2800$ shows the formation of an $A$-rich wetting layer 
at the surface ($z=0$) in conjunction with phase separation in the bulk. This 
wetting layer propagates into the bulk, as seen from the depth profiles. The 
order parameter $\psi(\vec{r},t)$ is defined from the local densities as 
$\psi(\vec{r},t)=(n_A-n_B)/(n_A+n_B)$. The quantity $\psi_{\textrm{av}}(z,t)$ is 
obtained by averaging $\psi(\vec{r},t)$ in the $x,y$ directions (parallel to 
the surface), and further averaging over $50$ independent runs. In the bulk, 
the spinodal decomposition wave-vectors are randomly oriented -- the above 
procedure yields $\psi_{\textrm{av}} \simeq 0$. Near the surface, we see a 
structured morphology consisting of a wetting layer at the surface, depletion 
layer adjacent to it, etc. This layered structure propagates into the bulk. These 
SDSD profiles have been observed in many experiments on this problem 
\cite{jnkbw91,kdkb93,lwll09,wcl10,gk03}. 

\begin{figure}[!htbp]
\centering
\includegraphics*[width=0.95\textwidth]{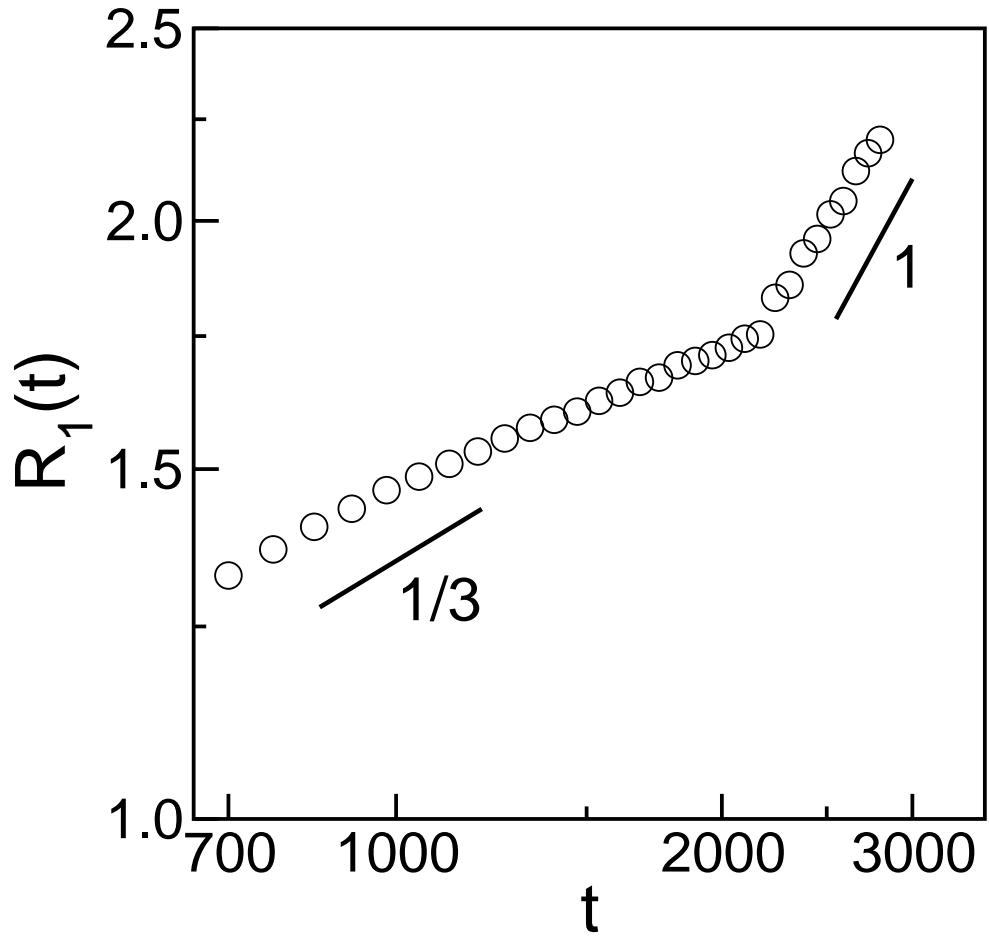}
\caption{Time-dependence of the wetting-layer thickness [$R_1(t)$ vs. $t$] of 
the SDSD profiles on a log-log scale. The straight lines have slope 1/3 
(diffusive regime) and 1 (viscous hydrodynamic regime), respectively.}
\label{fig:fig2}
\end{figure}

An important characteristic of the SDSD profiles is the time-dependence of the 
wetting-layer thickness $R_1(t)$. This is defined as the first zero-crossing of
the laterally-averaged depth profiles in Fig.~\ref{fig:fig1}. In Fig.~\ref{fig:fig2},
we plot $R_1(t)$ vs. $t$ for the evolution shown in Fig.~\ref{fig:fig1}. The growth dynamics is 
power-law, $R_1(t) \sim t^\theta$, with $\theta \simeq 1/3$ for $t \lesssim 2000$ and 
$\theta \simeq 1$ for $t \gtrsim 2000$. For these system sizes, we can go up to 
$t \simeq 3000$ before encountering finite-size effects due to the lateral 
domain size becoming an appreciable fraction of the system size $L$. 

How can one understand this crossover? At early times, the wetting layer grows 
by the diffusive transport of $A$ from bulk domains of size $R\sim{(\sigma t)}^{1/3}$ 
(with chemical potential $\mu \simeq \sigma/R$, $\sigma$ being the surface tension) 
to the flat surface layer of size $\simeq \infty$ (with $\mu \simeq 0$). If we 
neglect the very early potential-dependent growth regime 
\cite{pb01}, 
we have 
\begin{equation}
 \frac{dR_1}{dt} \simeq \frac{\sigma}{R h} \simeq \frac{\sigma}{R R_1},
 \label{eq3:R1t}
\end{equation}
where $h \sim R_1$ is the thickness of the depletion layer (see Fig.~\ref{fig:fig1}). 
From Eq.~(\ref{eq3:R1t}), 
we readily obtain the Lifshitz-Slyozov (LS) growth law: $R_1 \sim {(\sigma t)}^{1/3}$. 
At later times, bulk tubes establish contact with the wetting layer and material is 
pumped hydrodynamically to the surface. The subsequent growth is analogous to that 
in phase separation of fluids -- we expect $R_1(t) \sim t$ (viscous hydrodynamic 
regime) which crosses over to $R_1(t) \sim t^{2/3}$ (inertial hydrodynamic regime). 
The latter stage is presently not accessible via MD simulations \cite{adp10}, 
due to computational limitations. However, our results for wetting-layer dynamics 
in Fig.~\ref{fig:fig2} appear to access the viscous regime, albeit in a limited 
time-window. We remark that the crossover time ($t_c \simeq 2000$) is consistent
with that reported by Ahmad et al. \cite{adp10} in an MD simulation of bulk phase
separation with similar parameter values. Clearly, we need substantially larger
system sizes to obtain an extended regime of linear growth in the viscous regime.
Notice that the crossover in Fig.~\ref{fig:fig2} is quite sharp, suggesting that there
is a rapid pumping of material to the wetting layer when the bulk tubes first make contact.

\begin{figure}[!htbp]
\centering
\includegraphics*[width=0.95\textwidth]{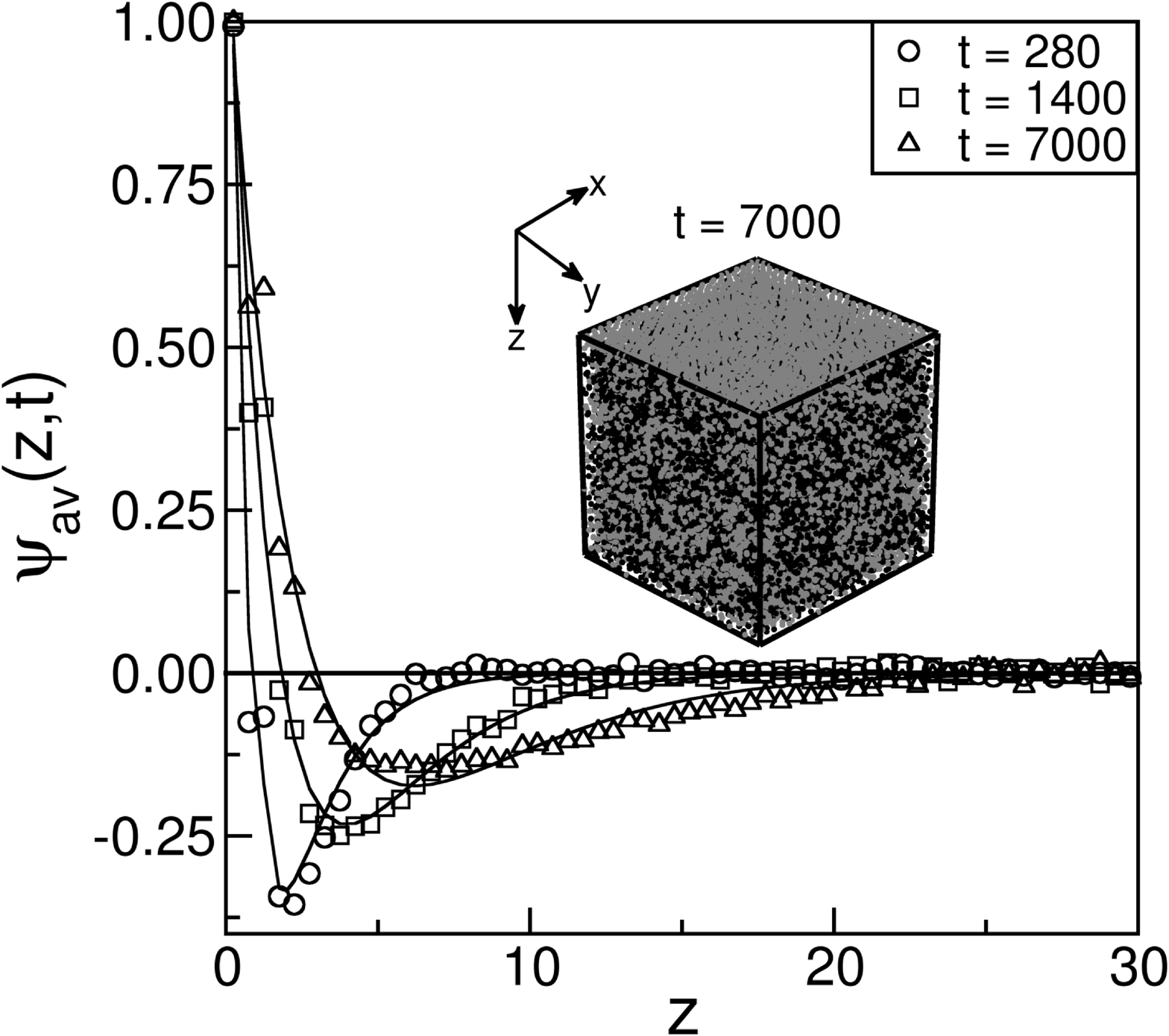}
\caption{Kinetics of surface enrichment (SE) for an $AB$ mixture in contact with 
a surface (located at $z=0$) which attracts the $A$-particles. We set 
$\epsilon_r=0.5,\epsilon_a = 3.0$, and $T=2.0\simeq 1.41\,T_c$. 
The other simulation details are provided in the text. Main figure: Plot of 
$\psi_{\textrm{av}}(z,t)$ vs. $z$ at $t=280, 1400, 7000$ MD units. 
The double-exponential fits for the SE profiles are shown as solid lines. 
Inset: Evolution snapshot at $t=7000$. The $A$-particles are marked gray and 
the $B$-particles are marked black.}
\label{fig:fig3}
\end{figure}

Second, we present results for the kinetics of SE. In this case, the system size was 
$L^2 \times D$ with $L=32, D=64$ ($N=66536$). As the bulk remains homogeneous,
 the lateral size $L$ (in the $x, y$ directions) is not severely constrained. However, 
in the direction perpendicular to the surface at $z=0$, we need sufficiently 
large $D$ to ensure decay of the enrichment profiles as $z \to D$. For the range of
times studied here ($t \leq 7000$), test runs with other linear dimensions showed 
that $D=64$ is large enough to eliminate finite-size effects. In Fig.~\ref{fig:fig3}, 
we show the laterally-averaged profiles and an evolution snapshot (inset) for the 
kinetics of SE. The quench temperature was $T=2.0 \simeq 1.41 T_c$. The depth profiles 
were obtained using the same procedure as for SDSD, as an average over $50$ independent 
runs. The surface potential parameters were $\epsilon_r=0.5$ and $\epsilon_a=3.0$. 

As expected, the morphology for SE is quite different from that in the SDSD case 
(cf. Fig.~\ref{fig:fig1}). There is a thin enrichment layer of $A$ at the surface. 
Due to the conservation of the order parameter, there must be a corresponding 
depletion layer which decays to $\psi_{\textrm{av}} \simeq 0$ in the bulk. These 
profiles are in agreement with the experimental observations of Jones et al. 
\cite{jkrss89} 
on blends of deuterated and protonated polystyrene, and the experimental results 
of Mouritsen 
\cite{mouritsen} 
on biopolymer mixtures. Notice that similar profiles are seen for SDSD if the system 
is quenched to the metastable region of the phase diagram 
\cite{pb01}. 
The evolution dynamics in that case is analogous to the SE problem, as long as 
droplets are not nucleated in the system. 

For the case with diffusive dynamics, Binder and Frisch \cite{bf91} and Frisch et al. 
\cite{fpn99,pf93} have studied the morphology of SE profiles in the framework of a
linear theory. They find that the SE profiles have a double-exponential form:
\begin{eqnarray}
 \psi(z,t)\simeq B_-(t)\,\mathrm{e}^{-z/\xi_-(t)}-B_+(t)\,\mathrm{e}^{-z/\xi_+(t)}, 
 \label{eq4:expfit}
\end{eqnarray}
with amplitudes $B_-(t), B_+(t)>0$. The quantities $B_-(t)$ and $\xi_-(t)$ 
rapidly saturate to their equilibrium values $a_1$ and $b_1$, which depend on the 
surface potential. The other length scale $\xi_+(t)$ grows diffusively with time, 
and $B_+(t)$ shows a corresponding decay:
\begin{equation}
 B_+(t) \simeq a_2\,t^{-1/2}, \qquad \xi_+(t) \simeq b_2\,t^{1/2}.
 \label{eq6:plus}
\end{equation}
The conservation constraint requires that $B_-\xi_-=B_+\xi_+$. In 
Fig.~\ref{fig:fig3}, we see that the double-exponential function in 
Eq.~(\ref{eq4:expfit}) describes the SE profiles very well. 

\begin{figure}[!htbp]
\centering
\includegraphics*[width=0.98\textwidth]{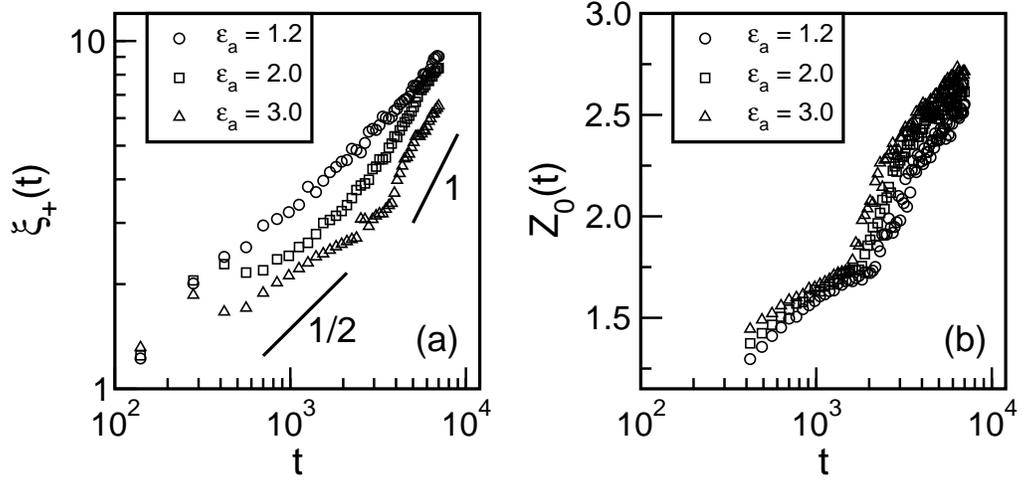}
\caption{Evolution kinetics of SE profiles. (a) Time-dependence of $\xi_+$ 
on a log-log scale for $\epsilon_r=0.5$ and $\epsilon_a = 1.2,2.0,3.0$. 
The lines of slope 1/2 and 1 correspond to diffusive and hydrodynamic 
growth, respectively. 
(b) Time-dependence of the zero-crossing $Z_0(t)$ on a log-linear scale.}
\label{fig:fig4}
\end{figure}

What about the parameters of the double-exponential profile? As for the diffusive case,
we find that $B_-$ and $\xi_-$ rapidly saturate to their equilibrium values and may be 
treated as static quantities. The growing length scale is $\xi_+(t)$ -- we plot 
$\xi_+$ vs. $t$ for different surface field strengths in Fig.~\ref{fig:fig4}(a). 
The early-time dynamics is consistent with diffusive growth ($\xi_+ \sim t^{1/2}$), 
but again there is a crossover at $t\sim t_c$ to a hydrodynamic regime ($\xi_+ \sim t$). 
The crossover is also seen in the thickness of the SE layer. For the 
profile in Eq.~(\ref{eq4:expfit}), the zero is located at $Z_0(t)\simeq \xi_- 
\ln(B_-/B_+)$. Thus, we expect $Z_0(t)\sim\ln t$ in both the time-regimes of 
Fig.~\ref{fig:fig4}(a), but the slope should be steeper for $t>t_c$. This is 
precisely the behavior seen in Fig.~\ref{fig:fig4}(b), where we plot our MD 
results for $Z_0(t)$ vs. $t$ on a log-linear scale. This confirms that there 
is a crossover in the growth exponent from the diffusive regime 
($\theta \simeq 0.5$) to the hydrodynamic regime ($\theta \simeq 1.0$).

To understand the crossover in Fig.~\ref{fig:fig4}, consider the dimensionless 
evolution equation for the order parameter in the presence of a fluid velocity 
field $\vec{v}(\vec{r},t)$ 
\cite{bray94,pw09}:
\begin{equation}
 \frac{\partial}{\partial t}\psi(\vec{r},t) = \nabla^2 \mu 
-\vec{v}\cdot \vec{\nabla}\psi ,
 \label{eq7:psir}
\end{equation}
where the chemical potential (for $T>T_c$) is $\mu=\psi+\psi^3-(1/2)\nabla^2\psi$. 
For the SE problem, the system is homogeneous in the directions parallel to 
the surface. Thus, we set $\psi(\vec{r},t) \simeq \psi(z,t)$ and 
$v_z(\vec{r},t) \simeq v_z(z,t)$. Then, Eq.~(\ref{eq7:psir}) becomes 
\begin{equation}
 \frac{\partial}{\partial t}\psi(z,t) = \frac{\partial^2 \mu}{\partial z^2} 
- v_z \frac{\partial \psi}{\partial z} .
\label{eq8:psiz}
\end{equation}
We use the functional form of $\psi(z,t)$ in Eq.~(\ref{eq4:expfit}) to estimate 
the dominant contribution to the various terms in Eq.~(\ref{eq8:psiz}) at 
$z \sim O(\xi_+)$, i.e., far from the surface \cite{jpd10}. We have 
\begin{equation}
 \frac{\partial \psi}{\partial t} \sim \frac{1}{\xi_+^2}\frac{d\xi_+}{dt}, \qquad
 \frac{\partial^2 \mu}{\partial z^2} \sim \frac{1}{\xi_+^3}, \qquad
 v_z \frac{\partial \psi}{\partial z} \sim \frac{v_z}{\xi_+^2}. \label{eq9:terms}
\end{equation}

For $T>T_c$, the bulk is homogeneous and there is no large-scale structure 
formation in the composition or velocity fields. As the system is incompressible, the
velocity field obeys $\vec{\nabla} \cdot \vec{v} = 0$ or $\partial v_z/\partial z
= 0$ in the laterally homogeneous case, so that $v_z \sim$ constant. Notice
that $v_z < 0$ as there is a net current of the preferred component A towards the
enrichment layer. This current is dissipated at the surface by the formation of
locally inhomogeneous structures. At early times, the diffusive term in Eq.~(\ref{eq8:psiz})
dominates, yielding $\xi_+ \sim t^{1/2}$. At late times, the convective term in 
Eq.~(\ref{eq8:psiz}) is dominant, giving  a crossover to $\xi_+ \sim v_z t$. 
The precise dependence of the crossover time on various physical parameters can 
be estimated by considering the dimensional version of Eq.~(\ref{eq7:psir}) in 
conjunction with the Navier-Stokes equation for the velocity field 
\cite{bray94,pw09}. 

In summary, we have presented comprehensive MD results for the kinetics 
of SDSD and SE in binary mixtures at wetting surfaces. Both the SDSD wetting layer 
and the SE layer show a crossover from a diffusive regime (with $R_1 \sim t^{1/3}$ 
and $\xi_+ \sim t^{1/2}$) to a hydrodynamic regime (with $R_1 \sim t$ and 
$\xi_+ \sim t$). These crossovers can be understood by simple phenomenological 
arguments. Our MD results are of great experimental relevance as most studies 
of these problems are done on polymer or fluid mixtures. We hope that our 
results in this paper will provoke fresh experimental interest in this problem, 
and our theoretical results will be subjected to experimental confirmation.

\newpage

\newpage

\newpage

\newpage

\end{document}